\documentclass[journal]{IEEEtran}

\ifCLASSINFOpdf
\else
   \usepackage[dvips]{graphicx}
\fi
\usepackage{url}
\usepackage{caption}
\usepackage{graphicx}
\usepackage{cite}
\usepackage{amsmath,amssymb,amsfonts}
\usepackage{graphicx}
\usepackage{textcomp}
\usepackage{xcolor}
\usepackage[linesnumbered,ruled,vlined]{algorithm2e}
\begin{document}

\title{Energy-Efficient 3D Deployment of Aerial Access Points in a UAV Communication System \\}

\author{Nithin Babu, \IEEEmembership{Student Member, IEEE},  Constantinos B. Papadias, \IEEEmembership{Fellow, IEEE}, Petar Popovski,  \IEEEmembership{Fellow, IEEE}
\thanks{©2020 IEEE. Personal use of this material is permitted. Permission from IEEE must be obtained for all other uses, in any
current or future media, including reprinting/republishing this material for advertising or promotional purposes, creating new
collective works, for resale or redistribution to servers or lists, or reuse of any copyrighted component of this work in other
works.
This version of the work has been accepted for publication in the IEEE COMMUNICATIONS LETTERS. This work is supported by the project PAINLESS which has received funding from the European Union’s Horizon 2020 research and innovation programme under grant agreement No 812991. }
\thanks{N. Babu and C. B. Papadias are with Research, Technology and Innovation Network (RTIN), 
The American College of Greece, Greece (e-mail: nbabu@acg.edu, cpapadias@acg.edu).}
\thanks{N. Babu, C. B. Papadias and P. Popovski are with Department of Electronic Systems, Aalborg University, Denmark (e-mail: niba@es.aau.dk,cop@es.aau.dk,petarp@es.aau.dk)}}

\markboth{Accepted Version }
{Shell \MakeLowercase{\textit{et al.}}: Bare Demo of IEEEtran.cls for IEEE Journals}
\maketitle

\begin{abstract}
In this letter, we propose an energy-efficient 3-dimensional placement of multiple aerial access points (AAPs), in the desired area, acting as flying base stations for uplink communication from a set of ground user equipment (UE). The globally optimal energy-efficient vertical position of AAPs is derived analytically by considering the inter-cell interference and AAP energy consumption. The horizontal position of AAPs which maximize the packing density of the AAP coverage area are determined using a novel regular polygon-based AAP placement algorithm. We also determine the maximum number of non-interfering AAPs that can be placed in the desired area. The effect of the AAP energy consumption on the optimal placement and the analytic findings are verified via numerical simulations.
\end{abstract}
\IEEEpeerreviewmaketitle
\section{Introduction}
The aerial coverage provided for temporary data demand events with the help of unmanned aerial vehicles (UAVs) acting as flying base stations is considered as one of the essential components of fifth-generation (5G) and beyond-5G wireless networks. Unlike the conventional approach of fixed base stations, the portable feature of the UAV-based aerial communication system not only increases the probability of line-of-sight (LoS) links between the UEs and the AAP but also could be dynamically deployed in natural disaster areas\cite{DISASTER} or social events such as concerts. One of the major limitations of the UAV communication system is it's limited lifetime proportional to the available onboard energy. So the UAVs should be deployed in such a way as to increase the number of bits successfully transmitted per joule of energy consumed, defined as global energy efficiency (GEE). The GEE of the UAV-based system depends on the 3-D coordinates of the UAV location; as the altitude increases, the coverage area of the UAV increases and the UAV energy consumption increases \cite{energyquad}, thereby affecting the GEE. The horizontal positioning of the UAVs determines the fraction of the total number of users in the desired area covered by the UAV; the higher the fraction, the higher the GEE. The authors of \cite{alzenad20173} propose an energy-efficient 3-D placement of an unmanned aerial vehicle base station for  maximal coverage under the orthogonal multiple access (OMA) scheme. The work in \cite{mozaffari2016efficient} proposes an optimal 3-D deployment of three UAV-base stations in a given urban area for maximum coverage under the OMA scheme. In \cite{8269064}, an online method for proper 3D deployment of UAV base stations to maximize the lifetime of the network is proposed. None of the above works consider the energy consumption of the mechanical parts of the aerial vehicle and the co-channel interference from the neighboring cells. In our previous work \cite{Babu2008:Energy} we have determined the energy-efficient hovering altitude for a standalone AAP deployed for orthogonal downlink broadcast transmission. The energy-efficient 3-D deployment of multiple AAPs in a given geographical area considering both the communication-related energy and UAV energy consumption in the presence of inter-cell interference has, to the best of our knowledge, not been investigated in the literature.
 
In this letter, we analytically determine the optimal vertical position of AAPs by solving the GEE maximization of identical and independent single AAP systems with the altitude and the individual UE power constraints. Then the horizontal coordinates of the AAPs with non-overlapping coverage areas are determined by posing it as a problem of non-overlapping circle packing and solved using the proposed multilevel regular polygon-based placement algorithm.
\section{System Model}
We consider a circular geographical area of radius $R$, containing a set of $N_{u}$ uniformly distributed stationary ground UEs with a density $\rho_{u}$, such that $N_{u}=\rho_{u}\pi R^{2}$. As shown in Figure 1, the given geographical area is covered by multiple AAPs positioned in a way that their coverage areas do not overlap and the horizontal plane coordinates of the AAP are assumed to be those of the center of the AAP coverage area. A universal frequency reuse among the AAPs is assumed, in which the total bandwidth $W$ is equally divided among $N_{u}^{'}=\rho_{u}\pi R_{a}^{2}$ UEs lying in the circular coverage area of an AAP of radius $R_{a}$. Since the UEs in the neighboring cells use the same set of frequencies, the receiver at the AAP experiences inter-cell interference from the co-channel UEs in the neighboring cells. Let $\mathcal{U}^{'}$ be the subset of UEs covered by the AAP, such that $\vert{\mathcal{U}^{'}}\vert =\rho_{u} \pi R_{a}^{2}$.
\begin{figure}
\centering
\captionsetup{justification=centering}
\centerline{\includegraphics[width=0.9\columnwidth]{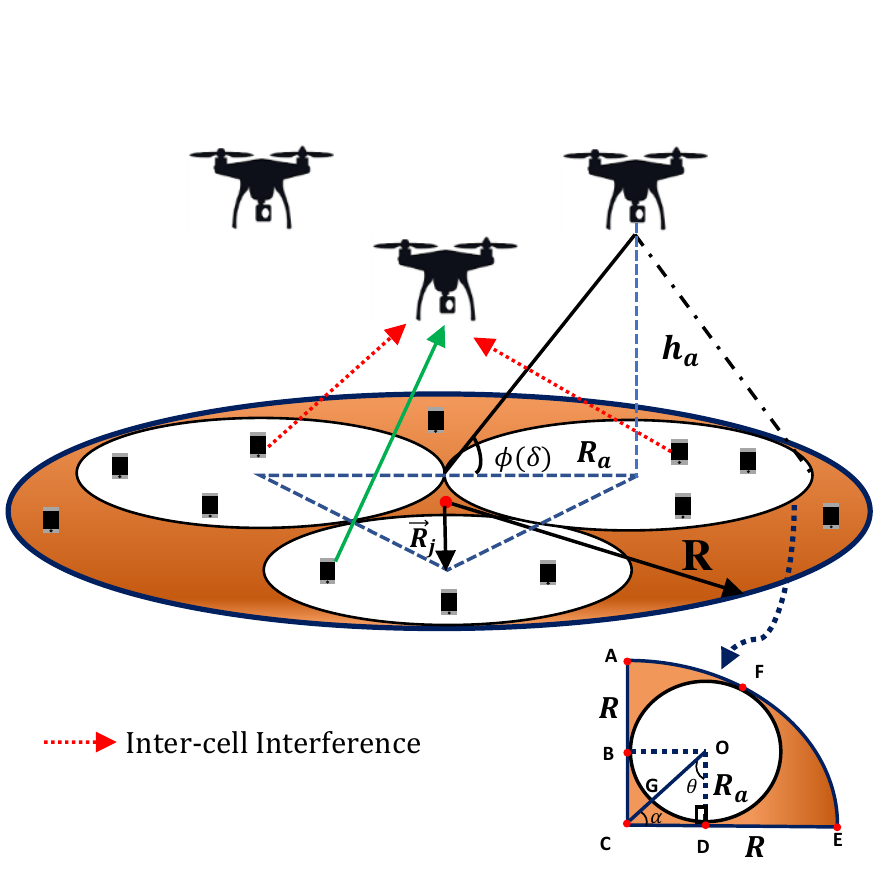}}
\caption{System setup.}
\label{figure1}
\end{figure}
\subsection{Optimal Vertical Positioning of the AAPs}\label{hopt}
Considering AAPs with non-overlapping coverage areas, the problem of finding the optimal vertical positioning of the AAPs breaks down to identical and independent single AAP vertical positioning problems. Hence all the AAPs will be hovering at the energy-efficient altitude obtained by solving the independent AAP vertical positioning problem \cite{8269064}. Here we consider an orthogonal uplink communication between the UEs and the associated AAP.  
\subsubsection{Channel model}
We consider the channel model proposed by authors in \cite{angleforlos}, in which the communication channel between the UEs and the AAP can be modelled either as a line-of-sight (LoS) or a non-line-of-sight (N-LoS) link. Since the planning phase of the base station deployment considers long-term channel variation rather than short-term random behavior, we neglect the small scale channel variations due to the dynamic propagation environment \cite{angleforlos}. The environment-dependent long-term channel variations due to shadowing and scattering referred to as additional path loss have a Gaussian distribution \cite{angleforlos}; however, in this letter, we only use the mean value of this distribution and not its random behavior \cite{alzenad20173}, \cite{8269064}, \cite{angleforlos}. Hence $\eta_{l}$ and $\eta_{nl}$ are the mean value of the additional path loss for LoS and N-LoS links. Then the path loss for the LoS and N-LoS links between a UE located at a distance of $r_{i}$ from the center of the coverage area is given by

\begin{IEEEeqnarray}{rCl}
L_{x} & = & {\frac{\eta_{x}d_{i}^{2} }{g_{0}}}\,\,\,\,\,\,\, \text{for}\,\,\, x\in\left\lbrace l, nl\right\rbrace   
\label{pl}
\end{IEEEeqnarray} where $g_{0}=(c/4\pi f_{c})^{2}$ represents the channel gain at a reference distance of 1m; $c, f_{c}$ are the velocity of light and carrier frequency of the radio signal; $d_{i}=\sqrt{r_{i}^{2}+h_{a}^{2}}$, is the distance between the $i^{th}$ UE and the AAP. The LoS probability, $P_{l}$ between a UE and it's associated AAP is given by \cite{angleforlos};

 \begin{IEEEeqnarray}{rCl}
P_{l} & = & \dfrac{1}{1+a \exp\left[-b(\phi_{i}-a))\right]}
\label{plos}
\end{IEEEeqnarray}
where $a,b$ are environment-dependent parameters given in \cite{angleforlos} and $\phi_{i}=(180/\pi)\text{tan}^{-1}(h_{a}/{r_{i}})$ is the elevation angle between the $i^{th}$ UE and associated AAP. Hence, the N-LoS link probability, $P_{nl}$ associated with the same UE-AAP pair is $1-P_{l}(r_{i},h_{a})$. Because of the non-availability of the terrain knowledge, we consider a probabilistic mean path loss given by
\begin{IEEEeqnarray}{rCl}
\overline{L}(r_{i},h_{a}) & = & {P_{l}\times L_{l}+P_{nl} \times L_{nl}}\nonumber\\
& = & \frac{d_{i}^{2} }{g_{0}} \times [\eta_{l}P_{l}+\eta_{nl}(1-P_{l})]\nonumber\\
& = & \underbrace{\frac{d_{i}^{2} }{g_{0}}}_{\text{FSPL}} \times \underbrace{[\eta_{nl}+P_{l}(\eta_{l}-\eta_{nl})]}_{\eta_{m},\,\,\text{mean\,\,additional\,\, path\,\, loss}}
\label{mpl}
\end{IEEEeqnarray}
\subsubsection{AAP coverage region}In this letter, we consider the GEE as the performance matrix for the AAP deployment. Since the GEE of the considered system might not be maximum at the altitude corresponding to the minimum required SNR value\cite{angleforlos}, \cite{alzenad20173}, we define the coverage region of an AAP based on the $P_{l}$ threshold, $\delta$. For a given AAP altitude, all the UEs having a LoS probability greater than $\delta$ are considered to be covered by the AAP. This threshold translates into a circular coverage region with radius, $R_{a}=h_{a}\text{cot}(\phi(\delta))$ and all the UEs at distance $r_{i}\leq R_{a}$ are considered to be lying in the AAP coverage area.
\subsubsection{Uplink Power Control}
Each user chooses the transmit power according to the uplink power control specified in the 3GPP technical report \cite{3gpppc}. Then the transmit power for the $i^{th}$ UE (in Watts) is given by 
\begin{IEEEeqnarray}{rCl}
P_{i} &=& {\text{min}\left\lbrace P_{\text{max}}, P_{a}B(\overline{L}(r_{i},h_{a}))^\beta\right\rbrace}
\label{pi}
\end{IEEEeqnarray}
where $P_{\text{max}}$ is the maximum transmit power; $ P_{a}$ is the target arrived power at the AAP; $B$ is the number of allocated resource blocks and $\beta$ is the path loss attenuation factor of fractional transmission power control (TPC) \cite{3gpppc} \cite{zhang2016uplink}. The information about the target power and the AAP location is sent to the UEs through the control signaling. Since $B$ and $\beta$ do not depend on the AAP position, we assume they are both equal to 1. However, the algorithm developed in Section \ref{hp} is applicable for any $B,\, \beta$ values. 
Hence the average transmit power transmitted by the $i^{th}$ UE is given by 
 \begin{IEEEeqnarray}{rCl}
\overline{P}_{i} &=&  {P_{a}\overline{L}(r_{i},h_{a})} \,\,\,\,\,\,\,\forall\,\, i \,\,\in \mathcal{U}^{'}
\label{api}
\end{IEEEeqnarray}
Thus the expectation of the sum of the powers transmitted by all the UEs in the AAP coverage area is obtained by taking an expectation over the uniformly distributed UEs with a density $\rho_{u}$:

\begin{IEEEeqnarray}{rCl}
\overline{P}_{t} &=& { \rho_{u} \int_{0}^{R_{a}} 2 \pi P_{a}\overline{L}(r_{i},h_{a})r_{i}dr_{i}}\label{6}\nonumber \\
 & \leq & {\dfrac{2\pi \rho_{u} P_{a} \eta_{m} \text{cot}^{2}(\phi(\delta))h_{a}^{4}( \text{cot}^{2}(\phi(\delta))+2)}{4 g_{0}}}\label{7}
\label{apt}
\end{IEEEeqnarray}
The free space path loss (FSPL) and the $P_{l}$ variable of $\eta_{m}$, of $\overline{L}(r_{i},h_{a})$ depend on $r_{i}$. By \eqref{plos}, $P_{l}(r_{j})\geq P_{l}(r_{k})$ for all $r_{j}\leq r_{i}$. Because of the complex $P_{l}$ expression, for the remaining analytical derivation, we approximate $P_{l}$ for all the UEs in the AAP coverage area to be equal to the $P_{l}$ of the edge UE ($r_{i}=R_{a}$). Assuming no interference cancellation techniques at the AAP surrounded by $M$ AAPs, the upper bound of the data rate of the $i^{th}$ UE in bits per seconds (bps) is given by
\begin{IEEEeqnarray}{rCl}
D_{i} &=& {W_{i} \text{log}_{2}\left\lbrace 1 +\dfrac{\dfrac{\overline{P}_{i}}{\overline{L}(r_{i},h_{a})}}{\Sigma_{j=1}^{M}\dfrac{\overline{P}_{i,j}}{\overline{L}(r_{i,j},h_{a})}+\dfrac{\sigma_{0}^{2}W}{N_{u}^{'}}}\right\rbrace}\label{8}\\
 & = & {W_{i}\text{log}_{2}\left\lbrace 1 +\dfrac{P_{a}N_{u^{'}}}{MP_{a}N_{u^{'}}+\sigma_{0}^{2}W}\right\rbrace} \label{Di}
\end{IEEEeqnarray}
where $\sigma_{o}^{2}$ is the power spectral density of the zero-mean additive white Gaussian noise at the corresponding receiver; $W_{i}=W/N_{u^{'}}$. Since the inter-cell interference is a decreasing function of the distance from the receiver, in \eqref{Di}, we consider the case of maximum interference from the co-channel UEs in the neighboring cells lying close to the cell edge UEs. Because of the uplink power control, all the UEs in the coverage region will have the same data rate upper bounded by \eqref{Di}.  Assuming optimal (capacity-achieving) coding, we consider that these bounds will be attained. Then, the sum of the data rate will be: 
\begin{IEEEeqnarray}{rCl}
D_{u^{'}}(h_{a})& =&  \underbrace{\rho_{u} \pi h_{a}^{2}\text{cot}^{2}\phi(\delta)}_{N_{u^{'}}}\times D_{i}\label{su}
\end{IEEEeqnarray}
The sum of the data rates of the ground UEs, the data transmission energy, and the AAP energy consumption are the three major factors affecting the GEE of the system. The GEE of the considered system is defined as; 
\begin{IEEEeqnarray}{c}
\text{GEE}(h_{a})=\dfrac{S_{u^{'}}(h_{a})}{E(h_{a})}
\label{gee}
\end{IEEEeqnarray}
where ${S_{u^{'}}}(h_{a})= T D_{u^{'}}(h_{a}) $ is the total number of data bits transmitted by the UEs in the AAP coverage area in $T$ seconds;  $E(h_{a})$ is the total energy consumed by the AAP. The energy consumed by the AAP is the sum of the energy required for data communication and the energy consumed by the mechanical parts of the UAV during climbing and hovering, and is given by:
\begin{IEEEeqnarray}{rCl}
E(h_{a})& = &{\underbrace{\left(\alpha_{cl}h_{a}+\beta_{cl}\right)}_{E_{a,\text{climb}}}+\underbrace{\left(\alpha_{ho}h_{a}+\beta_{ho}\right)T}_{E_{a,\text{hover}}}+\underbrace{\overline{P}_{D}T}_{E_{\text{data}}}}
\label{energy}
\end{IEEEeqnarray}
where $\overline{P}_{D}=\overline{P}_{t}+P_{C}$ is the total data communication power, with $P_{C}$ being the total hardware circuit power consumption and where $\overline{P}_{t}$ is given by \eqref{apt}. $\alpha_{cl},\beta_{cl},\alpha_{ho},\beta_{ho}$ are the constants related to the UAV \cite{energyquad}. The aerial vehicle's energy consumption, $E_{a}(h_{a})$ increases with an increase in the altitude, because the reduced air pressure at higher altitudes demands the generation of an additional force by the propeller of the aerial vehicle, which results in increased energy consumption \cite{energyquad}\cite{sohail2019energy}.
The problem of determining the optimal hovering altitude of the AAP which maximizes the GEE while satisfying the altitude and the individual UE power constraints can be formulated as:
\begin{IEEEeqnarray}{rCl}
\text{(P1)} & : &{ \underset{h_{A}}{\text{maximize}}\,\,\,\,   \text{GEE}(h_{a})}\nonumber\label{p1}\\
\text{s.t.} & &{ h_{\text{min}}\leq h_{a}\leq h_{\text{max}}}\label{c1}\\
 & &{P_{i} \, \leq\, P_{max}}\label{c21}
\end{IEEEeqnarray}
where $P_{\text{max}}$ is the maximum power available at each UE, $h_{\text{min}}$ and $h_{\text{max}}$ are the minimum and maximum permitted AAP altitude specified
by the aviation regulatory board respectively. \eqref{c21} can be equivalently translated into the altitude constraint, $h_{a}\leq h_{\text{max}}^{'}= \sqrt{\dfrac{P_{\text{max}g_{0}}}{P_{a}\eta_{m}(1+ \text{cot}^{2}(\phi(\delta))) }}$.
(P1) is solved by using proposition 1.
\subsubsection*{Proposition 1}\label{pro}For a given $\rho_{u}$, $P_{\text{max}}$, $\delta$, $P_{a}$, $\text{GEE}(h_{a})$ is a decreasing function of the hovering altitude of the AAP. 
\subsubsection*{Proof}
To prove the decreasing nature of $\text{GEE}(h_{a})$, the numerator, $S_{u^{'}}(h_{a})$ and the denominator, $E(h_{a})$ should be a non-increasing and increasing function of $h_{a}$,  respectively. From \eqref{energy}, $E(h_{a})$ is an increasing function of $h_{a}$ and it remains to prove that $\frac{\mathrm{d} S_{u^{'}}(h_{a})}{\mathrm{d} h_{a}} \leq 0  \,\,   \forall \,h_{a} \in \left\lbrace h_{\text{min}},\text{min}(h_{\text{max}},h_{\text{max}}^{'})\right\rbrace$, which is shown below:
\begin{IEEEeqnarray}{rCl}
\dfrac{\dfrac{\mathrm{d} S_{u^{'}}(h_{a})}{\mathrm{d} h_{a}}}{T W} & = &\dfrac{2\kappa(\phi(\delta))h_{a} \text{log}_{2}e}{\kappa(\phi(\delta))h_{a}^{2}+\dfrac{\sigma_{o}^{2}W}{M+1}}-\dfrac{2\kappa(\phi(\delta))h_{a} \text{log}_{2}e}{\kappa(\phi(\delta))h_{a}^{2}+\dfrac{\sigma_{o}^{2}W}{M}}\nonumber\\
\label{S}
\end{IEEEeqnarray}
where $\kappa(\phi(\delta))=P_{a}\rho_{u} \pi \text{cot}^{2}(\phi(\delta))$. From \eqref{S}, since $\dfrac{1}{T W}\dfrac{\mathrm{d} S_{u^{'}}(h_{a})}{\mathrm{d} h_{a}}\approx 0$ for $P_{a}>> \sigma_{o}^{2}W$, the numerator of the GEE is proved to be a non-increasing function of $h_{a}$. Hence according to Proposition 1, the solution of (P1), the optimal vertical position of AAPs for maximum GEE, is the minimum altitude $h_{\text{min}}$. Then the optimal LoS threshold value $\delta_{o}$ corresponding to $h_{\text{min}}$ is determined numerically in section \ref{res}. The corresponding radius of the individual AAP coverage region is $R_{a}= h_{\text{min}}\text{cot}(\phi(\delta_{o}))$. 
\subsection{Optimal Horizontal Positioning of the AAPs}\label{hp}
In this section, we aim to determine the optimal horizontal positioning of the AAPs in the given desired circular region of radius $R$ so that the packing density, defined as the ratio of area covered by the AAPs to the given desired area, is maximized. We consider an equal coverage region for all the AAPs with optimal radius $R_{a}= h_{\text{min}}\text{cot}(\phi(\delta_{o}))$.  We propose a multi-level regular polygon-based placement algorithm to determine the optimal horizontal positioning of the AAPs in the desired area. In the first level of Algorithm 1, $N_{a,1}$ AAPs with non-overlapping coverage areas are placed along the boundary of the desired area. In the next level, $N_{a,2}$  AAPs are placed along the boundary of the void circle, of radius $R_{2}$, formed at the center of the desired area after the first level arrangement. The $N_{a,l}$ value is determined by Proposition 2.
\subsubsection*{Proposition 2}\label{pro2}
The maximum number of non-overlapping circles, $N_{a, l}(\geq 3)$, of radius $R_{a}$ that can be placed along the boundary of a larger circle of radius $R_{l}$ should satisfy the following inequality:
\begin{IEEEeqnarray}{rCl}
N_{a,l}\left[\pi+\alpha(1+\text{sec}\theta)^{2}-\sqrt{3}(\dfrac{\pi+2\alpha}{\pi})-\theta\right] & \leq & {\dfrac{\pi R_{l}^{2}}{R_{a}^2}} 
\label{n}
\end{IEEEeqnarray}
where $\alpha=\left(\dfrac{\pi}{2}-\theta\right)$ and $\theta=\dfrac{\left(N_{a,l}-2\right)\pi}{2N_{a,l}}$ are the angles associated with the polygon whose vertices are the center of the AAPs coverage regions as marked in Figure \ref{figure1}.
\subsubsection*{Proof}\label{proof2} 
Consider Figure \ref{figure1}; the void around a circle along the boundary of the desired area is given by
\begin{IEEEeqnarray}{rCl}
\text{V}_{\text{Edge}}&=&{\text{A}_{\text{ABODEFA}}-\text{A}_{\text{BFDOB}}}\nonumber\\
&=& R_{a}^{2}\left[\alpha\left(1+\text{sec}\theta\right)^{2}-\text{tan}\theta-\dfrac{\sqrt{3}\left(\pi+2\alpha\right)}{\pi}\right]\label{vo1}
\end{IEEEeqnarray}
 in which $\text{A}_{\text{BFDOB}}$ is the space claimed by the sector BFDOB of angle $\pi+2\alpha$ \cite{gaspar2000upper}; the void at the center of the desired  area is given by
 \begin{IEEEeqnarray}{rCl}
\text{V}_{\text{center}}&=& \underbrace{R_{a}^{2}\text{tan}\theta}_{\text{A}_{\text{BODCB}}}-\underbrace{R_{a}^{2}\theta}_{\text{A}_{\text{BODGB}}}\label{v2}
\end{IEEEeqnarray}
The inequality \eqref{n} is based on the constraint that the sum of areas covered by $N_{a, l}$ AAPs and the void area should be less than the desired area; that is $N_{a,l}\left[\pi R_{a}^{2}+\text{V}_{\text{Edge}}+\text{V}_{\text{center}} \right]\leq {\pi R_{l}^{2}}$.
In each level of the AAP placement, the packing density maximization problem can be equivalently modeled as 

\begin{IEEEeqnarray}{rCl}
\text{(P2)}  & : & \underset{\vec{R}_{j}, j\in \left\lbrace 1,...,N_{a,l}\right\rbrace}{\text{maximize}}\,\,\,\,\dfrac{N_{a,l} R_{a}^{2}}{R_{l}^{2}}\label{m2}\\
\text{s.t.} & & \left \| \vec{R}_{j}-\vec{R}_{k} \right \|\geq 2R_{a}\,\,\,\forall\,\,j\neq k \in \left\lbrace 1,...,N_{a,l}\right\rbrace \label{c1}\\
& &\left \| \vec{R}_{j} \right \|+ R_{a}\leq R_{l}\,\,\,\,\,\,\,\forall\,\,j \in \left\lbrace 1,...,N_{a,l}\right\rbrace\label{c2} \\
& & {R_{l}\geq R_{a} = h_{\text{min}}\text{cot}[\phi(\delta_{o})]}\label{c3}
\end{IEEEeqnarray}
where $\vec{R}_{j}$ is the vector representing the location of the center of the coverage region of the $j^{th}$ AAP in the given geographical area. The maximum packing density is achieved when the AAPs coverage areas are non-overlapping and lie inside the desired area, and the voids between the coverage area are minimized.  The constraint \eqref{c1} guarantees the zero overlapping between the AAPs coverage areas; \eqref{c2} restricts the center of the AAP coverage region to be inside the void area. The constraint \eqref{c3} restricts the minimum geographical area to be covered greater than the coverage region of a single AAP. (P2) takes the form of a circle packing problem \cite{gaspar2000upper} and is solved using Algorithm 1. In Algorithm 1, \eqref{c1} is satisfied by placing the center of the inner circles of radius $R_{a}$ on the vertices of a regular polygon of $N_{a,l}$ sides of side length equal to $2R_{a}$ so that the tangency between the inner circles is achieved. The maximum value of $N_{a,l}$ satisfying \eqref{n} maximizes the objective function of (P2) while satisfying \eqref{c2}; $N_{a,l}\geq 3$ implies $R_{l}\geq R_{a}(1+\text{sec}30^{o})=2.1547R_{a}$.
 \begin{algorithm}[]
 \label{algo}
\caption{Multilevel regular polygon based AAP placement algorithm}
\textbf{Input}:\,$R_{a}$, $R$, $l=1$.\\
Find $N_{a,l}$ using \eqref{n} with $R_{l}=R$\\
 \While{$(1)$}
 {
 $l=l+1$\\
 \If{$\left\lbrace \left[R-2(l-1)R_{a}\right]\geq2.1547R_{a}\right\rbrace$}
 {
Find $N_{a,l}$ using \eqref{n} with $R_{l}=R-2(l-1)R_{a}$
 }  
 \Else
 {
break;
 }
 }
 \If{($R_{l}\geq R_{a}$) $\&$ ($R_{l} < 2R_{a}$)}
 {
$N_{a,l}=1$;
 }
 \If{($R_{l}\geq 2R_{a}$) $\&$ ($R_{l} < 2.1547R_{a}$)}
 {
$N_{a,l}=2$;
 }
 \textbf{Output}:\,{Obtain the horizontal coordinates of the AAP location using $N_{a,l}$ value} 
 \label{algorithm1}
\end{algorithm}
 In step 2 of Algorithm 1, the maximum number of non-interfering circles, $N_{a,1}$, that can be placed along the boundary of the desired area is determined using \eqref{n}. In step 5, if the void formed at the center of the desired area after the $l^{th}$ level circle arrangement contains a circle of radius $R-2lR_{a} \geq 2.1547R_{a}$, then in the next level,  $N_{a,l+1}$ circles can be placed in the center, where $N_{a,l+1}$ is determined using \eqref{n} with $R_{l+1}=R-2lR_{a}$. This multilevel circle packing continues until the maximum radius of the void circle at the center of the desired area is less than $2.1547R_{a}$. In the $l^{th}$ level, the coordinates of the horizontal location of the AAPs, which is same as the coordinates of the vertices of the regular polygon of $N_{a,l}$ sides can be obtained as $\left[R^{'}\text{cos}\left(\dfrac{2\pi m}{N_{a,l}}\right),R^{'}\text{sin}\left(\dfrac{2\pi m}{N_{a,l}}\right)\right]$ where $R^{'}=R-\dfrac{l(l+1)R_{a}}{2}$, $m\in \left\lbrace0, 1,..,N_{a,l}-1\right\rbrace$.   
\section{Simulation Result and Analysis}
\begin{figure}
\centering
\captionsetup{justification=centering}
\centerline{\includegraphics[width=0.9\columnwidth]{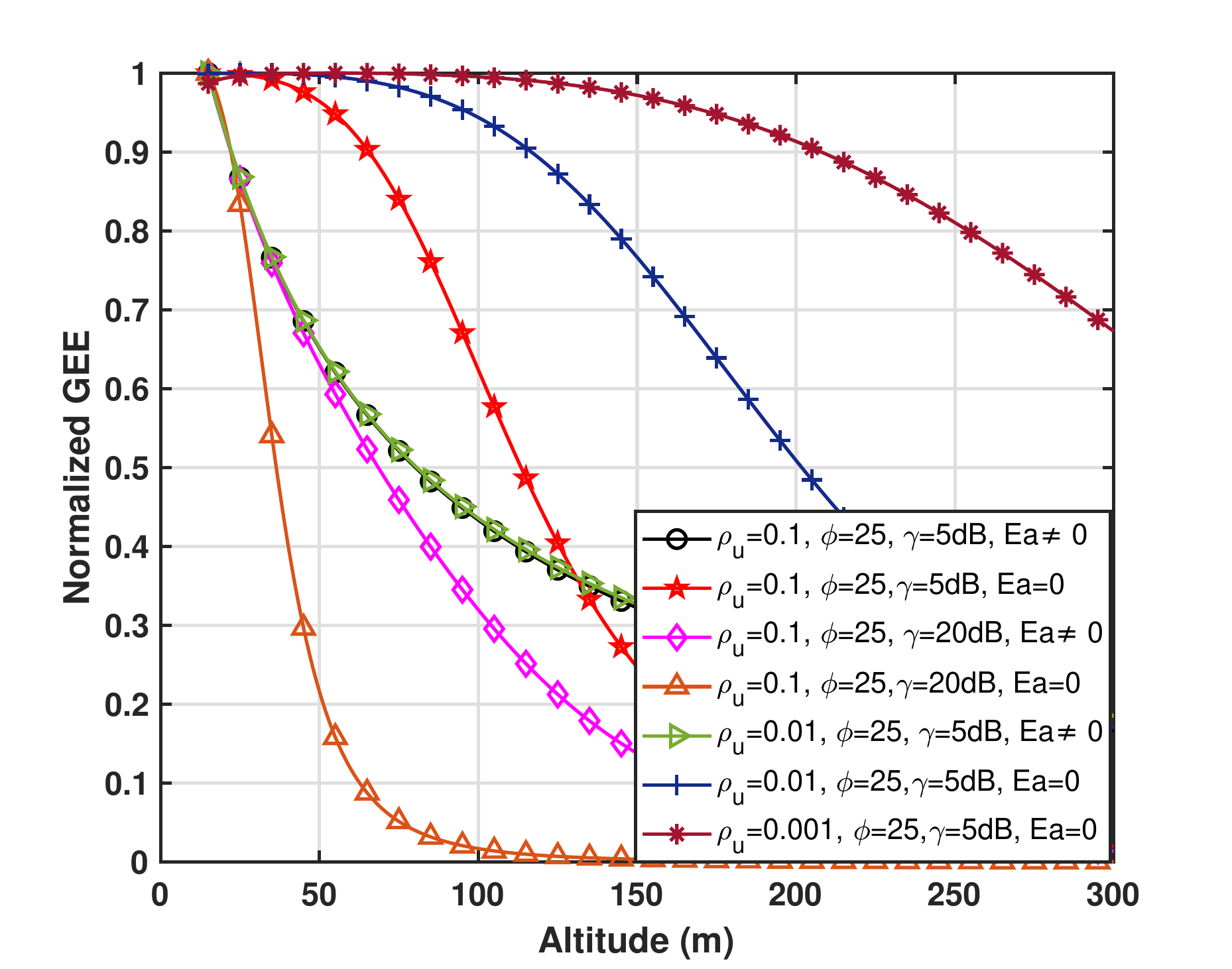}}
\caption{GEE variation with the hovering altitude.}
\label{figure2}
\end{figure}
\label{res}

In this section, we provide some representative simulation results in support of our analysis. The considered simulation parameters are $g_{0}=1.42\times10^{-4}, \eta_{l}=0.1\text{dB}, \eta_{nl}=21\text{dB}, a=4.88, b=0.43$\cite{angleforlos}, $ W=20\text{MHz}, M=6, P_{C}=5\text{W}, T=500\text{s}, P_{\text{max}}=1\text{mW}, h_{\text{max}}=300\text{m}, h_{\text{min}}=15\text{m}, \alpha_{cl}=315, \beta_{cl}=-211.261, \alpha_{ho}=4.917, \beta_{ho}=275.204$ \cite{energyquad}.
\begin{figure}
\centering
\captionsetup{justification=centering}
\centerline{\includegraphics[width=0.9\columnwidth]{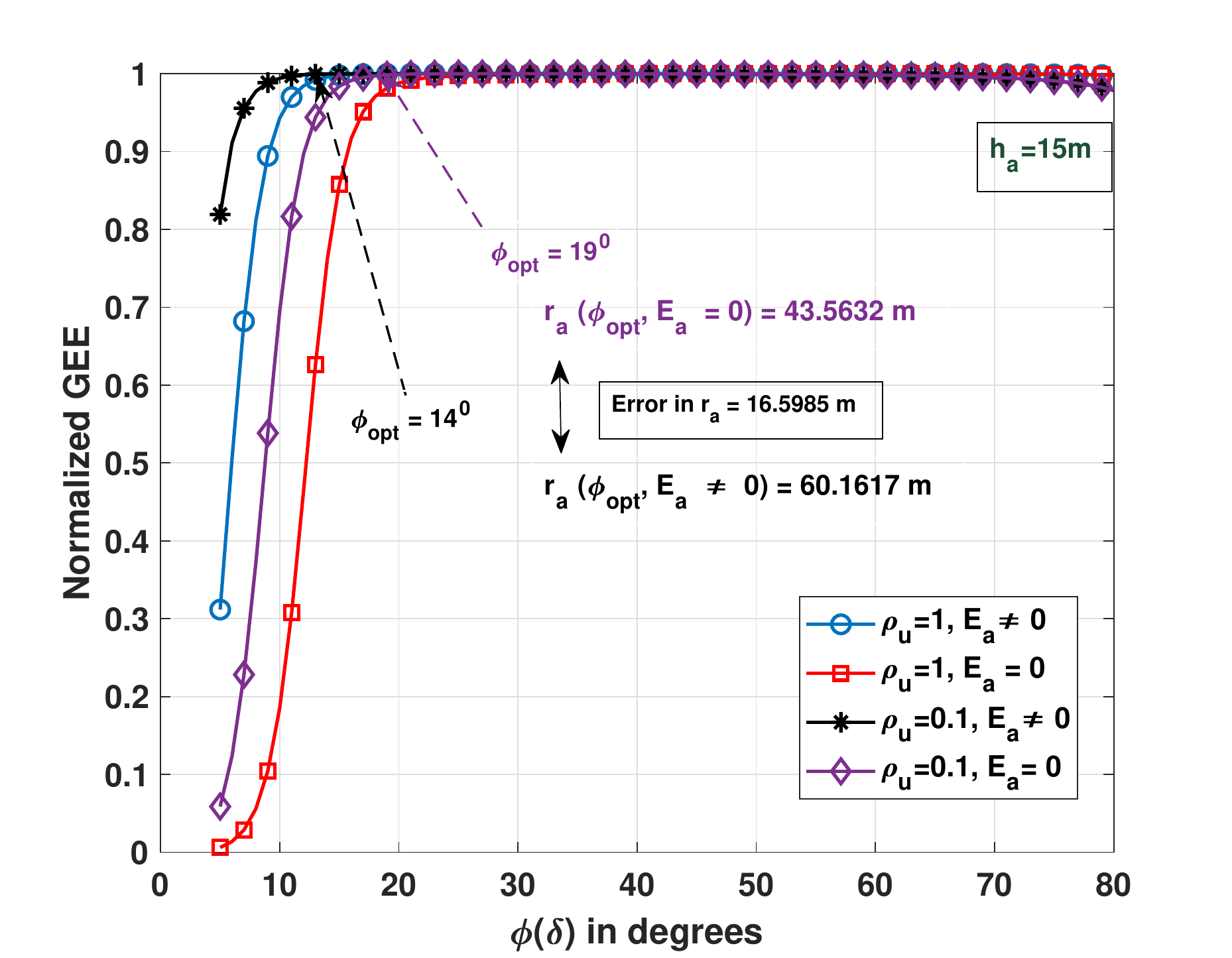}}
\caption{GEE variation for different $\phi(\delta)$ threshold.}
\label{figure3}
\end{figure}
Figure \ref{figure2} contains the plots of the GEE  with the hovering altitude of the AAP. The negative slope of the plots with non-zero AAP energy consumption ($E_{a}\neq 0$) verifies the monotonically decreasing nature of the GEE with the hovering altitude. This is because an increase in the number of UEs covered is highly compensated by  an increase in the communication-related and AAP energy consumption. As seen in Figure \ref{figure2}, in low signal-to-noise ratio ($\gamma=P_{a}/\sigma_{0}^{2}W$) regions, with the energy consumed by the aerial vehicle $E_{a}=0$, the GEE remains constant in low altitude region and then decreases leading to an error in determining the optimal hovering altitude. Since $E_{a}(h_{a})>> E_{data}$ in low UE density regions, the exclusion of aerial vehicle's energy consumption while defining the GEE will results in a non-optimal solution. On the other hand, with non-zero $E_{a}$, the GEE is a decreasing function of altitude. This explains the significance of $E_{a}$ in the energy-efficient placement of AAPs, a novel aspect of this letter.
\begin{figure}
\centering
\captionsetup{justification=centering}
\centerline{\includegraphics[width=0.9\columnwidth]{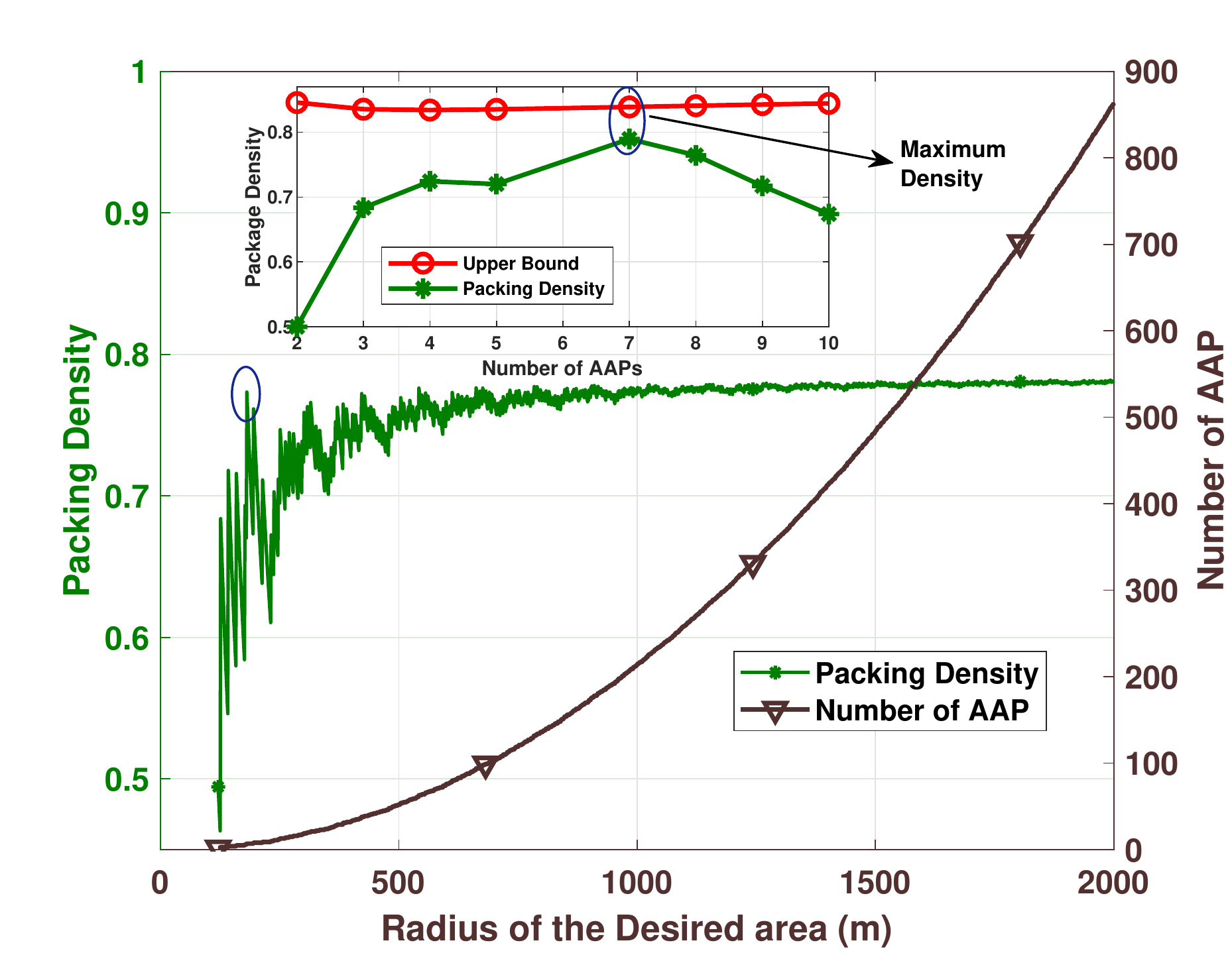}}
\caption{Packing Density for different desired area.}
\label{figure4}
\end{figure}
Figure \ref{figure3} shows the variation of the GEE with different $P_{l}$ threshold $\delta$, for a given AAP hovering altitude. Because of the saturation of $P_{l}$, and the proportional decrease in the number of covered UEs and the total transmit power, all the plots of Figure \ref{figure3} saturate after a particular $\phi(\delta)$ value. The saturation point shifts towards the left with non-zero $E_{a}$ value, because of the additional energy term in the denominator of the GEE. Figure \ref{figure4} gives the maximum packing density that can be achieved for the different radii of the desired area. It is observed that using Algorithm 1, for $R=180.48$m ($N_{a,1}=6, N_{a,2}=1$), a packing density almost equal to the Groemer’s upper bound on the maximum density of packing of $n$ equal circles in a circle \cite{gaspar2000upper} is achieved, For the remaining higher $R$ values, the AAPs placed using Algorithm 1 covers around 70 percent of the desired area. Figure \ref{figure4} shows the sample multi-level AAP placement pattern obtained through Algorithm 1 for two different values of the desired area. Considering the origin as the center of the desired area, for $R=180.48$m, the center coordinate set of the AAPs placed in the first level, forms the vertices of a regular hexagon and the next level contains a single AAP placed directly above the center of the desired region providing the packing density of 78.96\%. For $R=252.68$m, the first level of AAP placement follows an octagon, whereas the second level of AAP placement follows an equilateral triangle covering 68.44\% of the desired area. The packing density can be further improved by controlled overlapping between AAP coverage regions.
\section{Conclusion}
In this letter, we proposed the 3-D placement of a set of AAPs deployed for an energy-efficient uplink communication considering the inter-cell interference and AAP energy consumption. The energy-efficient hovering altitude of AAPs is analytically derived and the optimal horizontal positioning problem takes the form of a circle packing problem for maximum packing density, and solved using the multilevel regular polygon-based placement algorithm. The extension of our analysis to a downlink UAV-communication with non-uniformly distributed UEs along with the full coverage of the desired area by controlled overlapping between AAP coverage regions is left as future work. 
\begin{figure}
\centering
\captionsetup{justification=centering}
\centerline{\includegraphics[width=0.95\columnwidth]{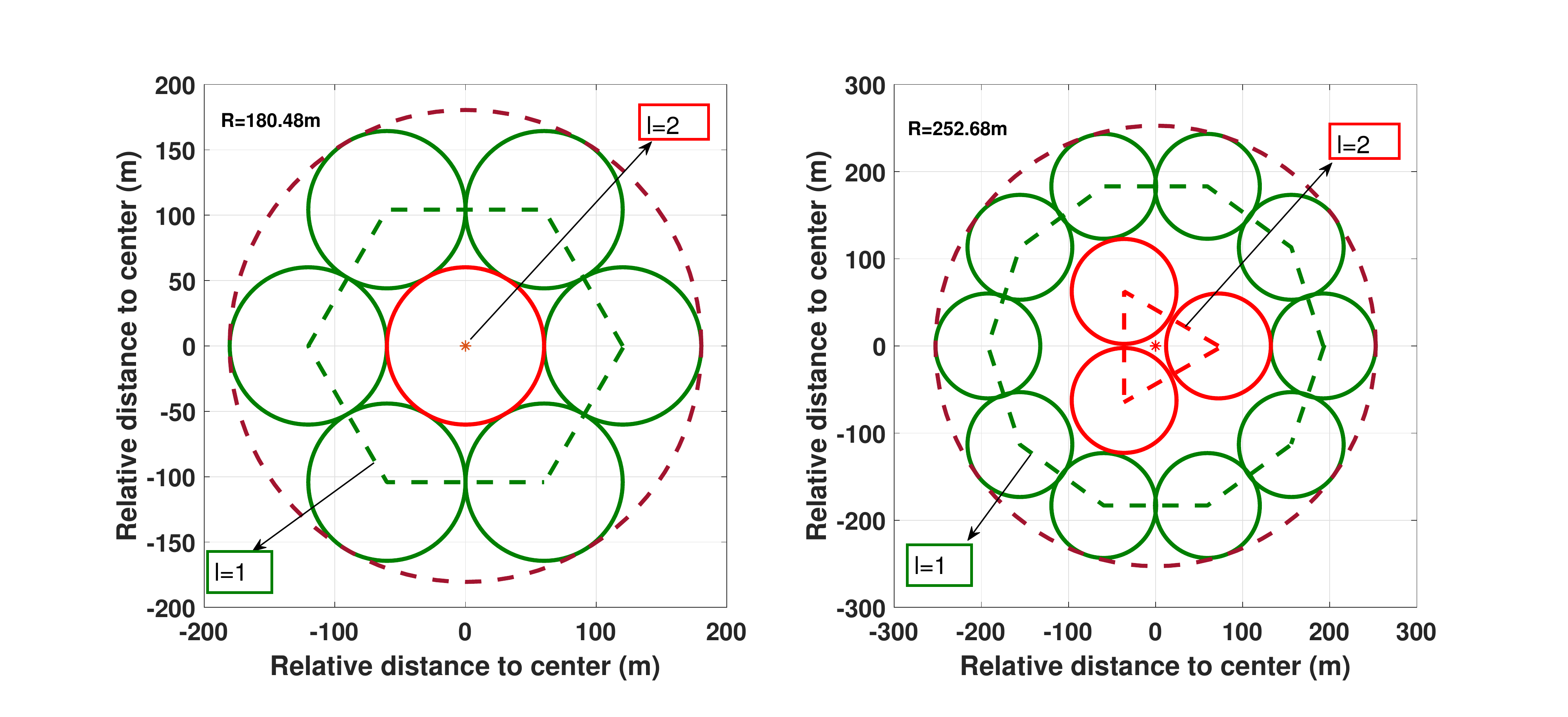}}
\caption{Horizontal Positioning of AAPs.}
\label{figure5}
\end{figure}
\bibliographystyle{IEEEtran}
\bibliography{IEEEabrv,ff}
\end{document}